\documentclass[conference]{IEEEtran}
\IEEEoverridecommandlockouts
\usepackage[english]{babel}
\usepackage[position=t,singlelinecheck=off]{subfig}
\usepackage{adjustbox}
\usepackage{algorithmic}
\usepackage{amsmath,amssymb,amsfonts}
\usepackage{booktabs}
\usepackage{cite}
\usepackage{color}
\usepackage{comment}
\usepackage{graphicx}
\usepackage{multirow}
\usepackage{subfloat}
\usepackage{textcomp}
\usepackage{tikz}
\usepackage{url}
\usepackage{xcolor,colortbl}
\def\BibTeX{{\rm B\kern-.05em{\sc i\kern-.025em b}\kern-.08em
    T\kern-.1667em\lower.7ex\hbox{E}\kern-.125emX}}
\usetikzlibrary{positioning}
\newcolumntype{L}[1]{>{\raggedright\let\newline\\\arraybackslash\hspace{0pt}}m{#1}}
\definecolor{sonialightgray}{rgb}{0.77,0.77,0.77}

\newcommand\copyrighttext{%
  \footnotesize \textcopyright 2012 IEEE. Personal use of this material is permitted. Permission from IEEE must be obtained for all other uses, in any current or future media, including reprinting/republishing this material for advertising or promotional purposes, creating new collective works, for resale or redistribution to servers or lists, or reuse of any copyrighted component of this work in other works. This paper has been accepted for publication in the 2020 Twelfth International Conference on Quality of Multimedia Experience (QoMEX).}
\newcommand\copyrightnotice{%
\begin{tikzpicture}[remember picture,overlay]
\node[anchor=south,yshift=10pt] at (current page.south) {\fbox{\parbox{\dimexpr\textwidth-\fboxsep-\fboxrule\relax}{\copyrighttext}}};
\end{tikzpicture}%
}

\begin{document}

\title{Impact of Tactile and Visual Feedback on Breathing Rhythm and User Experience in VR Exergaming
}

\author{
 \IEEEauthorblockN{Robert Greinacher$^1$, Tanja Koji\'c$^1$, Luis Meier$^1$,\\ Rudresha Gulaganjihalli Parameshappa$^1$, Sebastian M\"oller$^{1,2}$, Jan-Niklas Voigt-Antons$^{1,2}$}
 \IEEEauthorblockA{$^1$Quality and Usability Lab, TU Berlin, Germany\\
 $^2$German Research Center for Artificial Intelligence (DFKI), Berlin, Germany}
}


\maketitle
\copyrightnotice

\begin{abstract}
Combining interconnected wearables provides fascinating opportunities like augmenting exergaming with virtual coaches, feedback on the execution of sports activities, or how to improve on them. Breathing rhythm is a particularly interesting physiological dimension since it is easy and unobtrusive to measure and gained data provide valuable insights regarding the correct execution of movements, especially when analyzed together with additional movement data in real-time. In this work, we focus on indoor rowing since it is a popular sport that's often done alone without extensive instructions. We compare a visual breathing indication with haptic guidance in order for athletes to maintain a correct, efficient, and healthy breathing-movement-synchronicity (BMS) while working out. Also, user experience and acceptance of the different modalities were measured. The results show a positive and statistically significant impact of purely verbal instructions and purely tactile feedback on BMS and no significant impact of visual feedback. Interestingly, the subjective ratings indicate a strong preference for the visual modality and even an aversion for the haptic feedback, although objectively the performance benefited most from using the latter.
\end{abstract}



\section{INTRODUCTION}
A repetitive, monotonous movement is likely to affect the rhythm of breathing~\cite{bechbache1977entrainment}, which in return affects the exercise performance of the whole organism, e.g., ~\cite{durmic2015sport, ramacharaka2007science}. For adjusting the breathing pattern to a given exercise, the breathing-moving-synchronicity (BMS) is an interesting measure to quantify since it is cost-efficient and feasible to record mobile. Generally, wearables providing instant feedback are a promising research area. As a consequence of advances in technology, feedback systems are designed to provide specific information to assist during and after exercise. The appropriate feedback helps coaches and athletes to improve their performance and even accelerate recovery phases~\cite{broker2001advanced}. For creating automated feedback systems, the respiratory rate is an interesting measure for multiple reasons: first, it is tightly coupled with sportive performance, stress level, and the rhythm of a repetitive movement~\cite{gilbert2003clinical}. Also, it helps to adapt to stressful situations and to reduce stress levels \cite{gevirtz2000resonant}. Second, wearable sensors allow us to measure it even during movement, which makes it feasible to use as a dependent variable. Finally, breathing patterns can be communicated effectively in verbatim, visual animations, or tactile impulses~\cite{yu2015breathe}.

Although there are two known useful forms of breathing techniques for rowing, one is predominantly used for non-professional rowers: one exhale, and one inhale during one rowing stroke~\cite{steinacker1993pulmonary}. Since the ideal respiration is tightly coupled to the rowing strokes, the breathing pattern can be monitored and facilitated using an ergometer measuring the handlebar position and a breathing sensor worn by the athletes. The synchronicity between rowing and respiration, therefore, is an objective measure of how well the athlete maintained the correct breathing pattern. Furthermore, the ideal point in time can be derived to cue the athlete to breathe in or out.

Apart from optimizing breathing patterns of athletes during exercise, we want to couple the gained insights with another technology aiming to increase the attractiveness, experience, and success of personal training. The combination of sport exercise and gaming (exergaming), is defined as the activity of playing video games that are combined with physical movements that include strength, balance, and flexibility~\cite{oh2010defining}. The potential of VR exergaming lies in creating an effective training environment~\cite{kojic2019influence} and assisting training e.g., in the absence of coaches~\cite{ting2015application}. By using multisensory feedback combined with VR exergaming, we aim to make personal training healthier and more effective.

While audio-visual stimuli come with the potential disadvantaged of disturbing nearby users and possess privacy concerns \cite{janidarmian2016haptic}, haptic guidance is limited to the user wearing the interface. However, the literature still lacks evidence of a positive impact of the combination of haptic feedback for respiration together with VR exergaming. To our knowledge, this is the first study aiming to do so. We demonstrate how haptic impulses as feedback on the respiration rhythm during training helps to maintain a synchronous breathing-movement pattern and how a visual stimulation might be less suitable to do so - although the acceptance of participants indicates the contrary. Additionally, we present evidence for the independence of the VR exergaming environment on the effect of haptic feedback, suggesting the use of haptic feedback even outside the VR exergaming realm.

\subsection{Related work}
Sensory experience (SE) describes enhancing user Quality of Experience (QoE) of traditional audio-visual media systems by augmenting content with tactile or olfactory information, ambient light, or blowing air~\cite{timmerer2014sensory, waltl2010improving}. For SE of augmented audio and video content, a great body of research demonstrates increased QoE, e.g., ~\cite{giroux2019haptic, donley2014analysing}. The present paper aims to investigate SE in a VR environment, dynamically supporting a learning task. Pioneering work regarding learning in virtual environments suggests a positive impact of non-visual sensory cues on presence, spatial memory, and SE~\cite{ dinh1999evaluating}.

The literature agrees about the positive impact of multimodal feedback on motor learning, as outlined in the review by Sigrist et al.~\cite{sigrist2013augmented}. For example, haptic guidance~\cite{feygin2002haptic} helps novices to quickly learn movements in a safe and self-explanatory way~\cite{powell2012task}, while it can offer in-depth teaching aspects for advanced learners or athletes~\cite{sigrist2013augmented}. However, apart from position control feedback, the literature generally lacks haptic guidance investigations.

Multiple meta-analyses attested a favorable impact on reaction time and performance through multisensory feedback modalities~\cite{burke2006comparing,prewett2006benefits}. Both reviews concluded visual-tactile feedback decreases workload, which is in line with the Multiple Resource Theory~\cite{wickens2002multiple}, proposing different cognitive resources for processing different modalities of stimuli. These systems handle different chunks of information simultaneously. Hence, augmenting the rowing experience with an additional feedback modality should improve the rower's performance regarding BMS.

To increase the popularity of physical activity, motivation theory, which considers psychological needs and self-determination theory (SDT), has been investigated in the field of exergames by Peng et al.~\cite{peng2012need}. Results showed that with the introduction of different gamified elements, motivation, and engagement of players increased.  
Using a virtual environment as a medium for sports motivation has already been investigated with several sports activities such as biking~\cite{bolton2014paperdude} or rowing~\cite{kojic2018influence}. Using VR environments combined with rowing results in an increased feeling of flow and presence and even better overall user experience (UX)~\cite{schmidt2018impact}.




\section{METHODS}
\subsection{Participants}
Thirty-two participants took part in the study (12 female, aged 21-35 years, $M=27$, $SD=3.88$). Fifteen participants have had prior experiences with VR and stated they spend, on average, about 25.2min per week using it ($SD=1.09h$). We hypothesize the most significant impact of a technical intervention on supporting BMS in novices. Hence, a sample with little to no experience was chosen: 14 participants have used a rowing ergometer before, all without professional instructions. On average, they spent $7.8min$ per week ($SD=25.2min$). Following the Declaration of Helsinki the local ethics committee of the Faculty IV of the Technische Universität Berlin granted an ethics approval for this study (reference~FR\_2019\_01).

\subsection{Apparatus}
\begin{figure}[!htb]
  \centering
  \includegraphics[width=8.8cm]{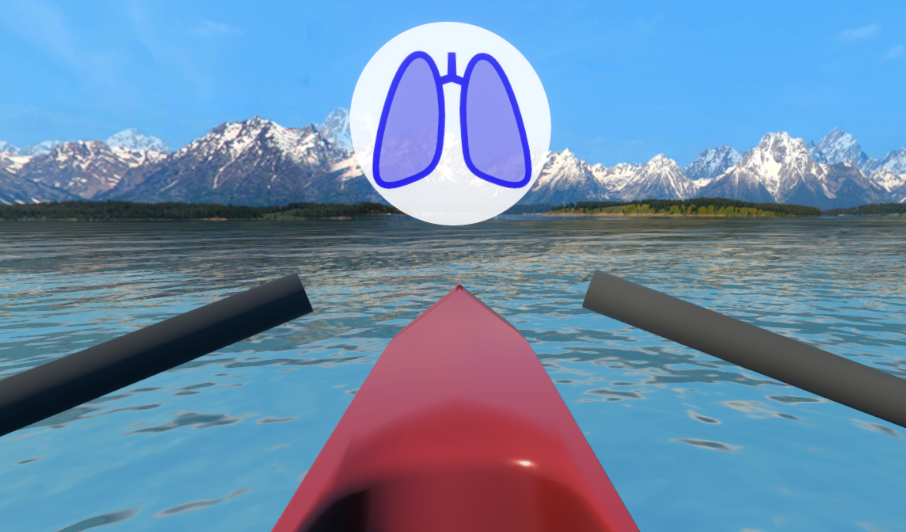}\\
  \caption{VR environment simulating a rowing boat. The lungs pictogram shows the fully inhaled state of the visual breathing feedback (the filling of the symbol moves up and down synchronous to the ideal chest expansion).}
  \label{fig:vrEnv}
\end{figure}

\begin{figure}[!htb]
  \centering
  \includegraphics[width=7cm]{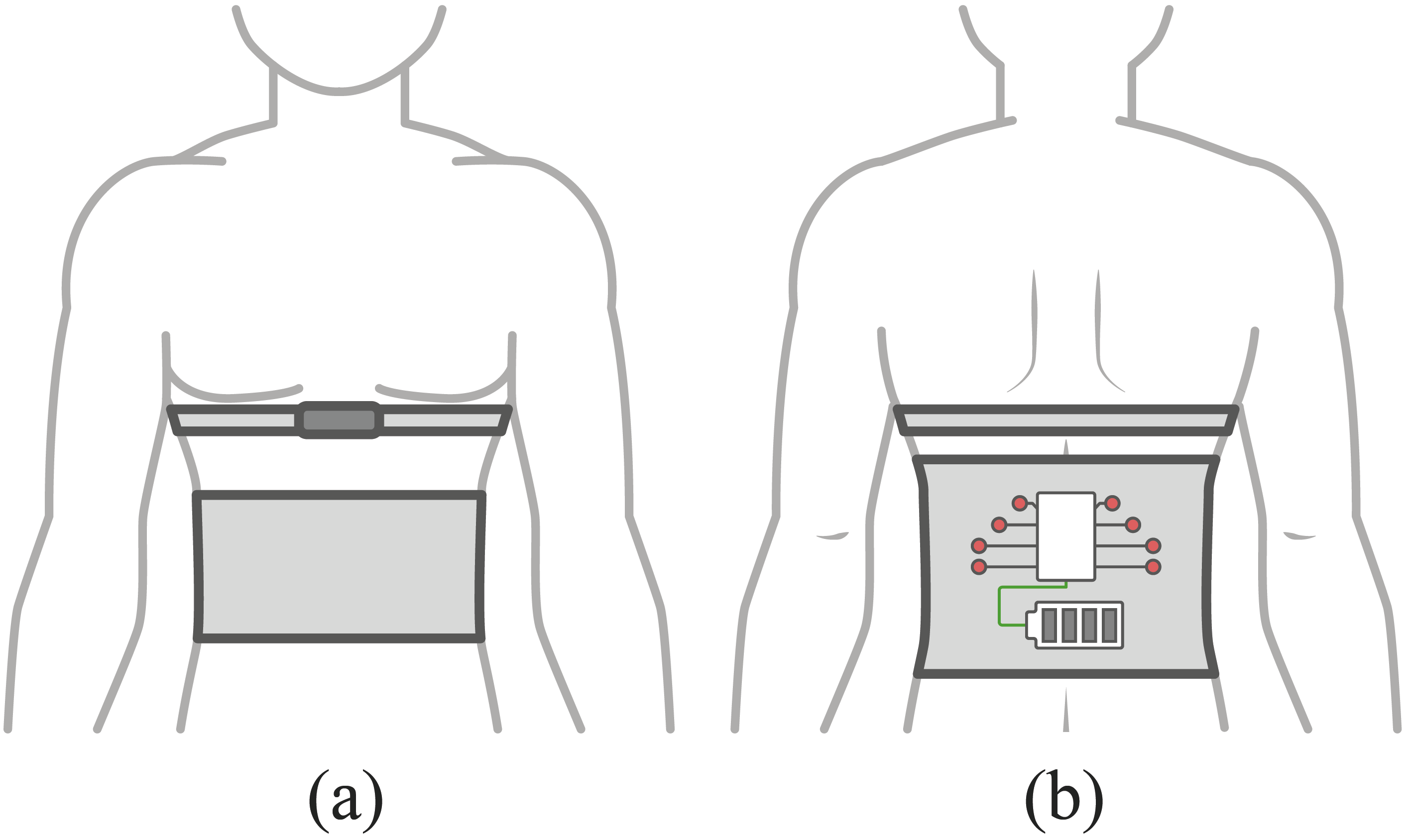}\\
  \caption{Breathing sensor and actuator setup, a) depicts the setup from the front, including the breathing sensor. b) shows the setup from the back together with the computer controlling the eight actuators used to provide haptic feedback.}
  \label{fig:FeedbackDesign}
\end{figure}

In order to compute synchronicity between breath and movement on the rowing machine, the setup involved a variety of components: As an ergometer, the indoor rowing machine \textit{Augletics Eight Pro}~\cite{augletics2019eight} was used. Its data was queried using a built-in REST API accessed over the integrated WiFi interface. To compare the participants' breathing rhythm while rowing, the SweetZpot FLOW~\cite{sweetzpot2019} breathing sensor was used to record the data. The participants were wearing the breathing sensor around the chest, as depicted in Figure~\ref{fig:FeedbackDesign}. The sensor measures expansion and contraction of inspiratory muscles and provides updates via Bluetooth at a rate of $10Hz$, each containing a chunk of seven records. The virtual environment was created in Unity and displayed on an HTC Vive~\cite{htcvive2019} head-mounted display (HMD). The environment consisted of a red boat with oars in the middle of a lake and, depending on the given condition, an animation of human lungs (Figure~\ref{fig:vrEnv}). The rowing oars moved synchronously with the rowing machine handle. As participants started rowing, the boat in the virtual environment propelled accordingly. The lungs pictogram provided visual breathing feedback using a gradually rising and declining virtual horizon, based on the participants' rowing movement and updated in real-time. The \textit{breathing actuator} provided by GHOST~-~feel~it.~GmbH provided a haptic impulse whenever athletes should start breathing out. This belt contained eight low-intensity vibration motors ($3.3V$, type ERM, arranged in a $4\times2$ matrix). Master electronics consisted of a µC, Bluetooth, and WiFi modules as well as a battery. The belt was strapped around the participants' waist, having the actuators stimulating the participants' lower back. Impulses were created through the simultaneous activation of all actuators for $200ms$. This feedback was timed right before the drive phase to stimulate exhalation. The orchestration of signals, environment, and actuators (via a REST API) was done in the Unity 3D application (see section \ref{OperationalizationSynchronicity} for more details). Apart from creating the lake scene \cite{schmidt2018impact}, Unity was used to collect the sensor data and to compute the correct timing for the next breathing impulse for the actuators given the current stretch of the lungs and the rowing handlebar position.

\subsection{Operationalization of Synchronicity}
\label{OperationalizationSynchronicity}
\begin{figure}[htb]
  \centering
  \includegraphics[width=8.8cm]{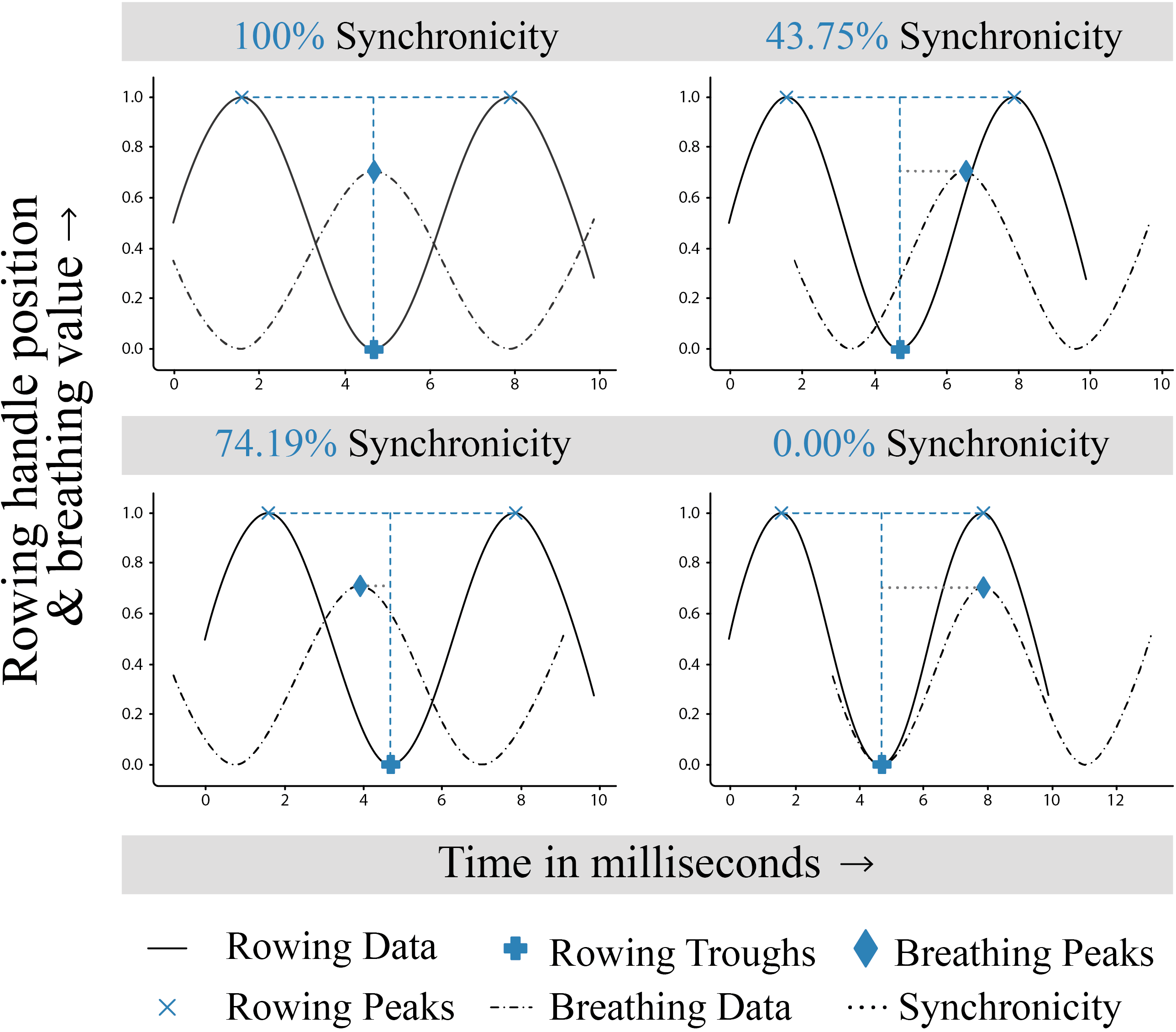}\\
  \caption{Synchronicity computation at the example of four single rowing cycles. Ideal synchronicity is achieved when the inhalation peak is at the center of two handle bar peaks while deviations cause the synchronicity to drop.}
  \label{fig:synchronicityCalculation}
\end{figure}

The handlebar position of the ergometer, together with respiration movements measured through chest expansion, is used to compute a synchronicity value for each rowing stroke (see Figure~\ref{fig:synchronicityCalculation}). The wave visualizing the handlebar movement consists of four phases: catch, drive, recovery, and the finishing phase. 

The rising part of the curve represents the drive phase in which participants are pulling the handle with their hands and are pushing themselves away from the starting position with their legs while keeping the back straight. Similarly, the falling part of the curve represents the recovery phase in which participants are leaning towards the front of the rowing machine and having the legs tucked. In Figure~\ref{fig:synchronicityCalculation}, the wave trough of the rowing data represents the catch phase. Likewise, the wave peak of the rowing data represents the finishing phase. Concerning breathing data, the rise represents the inhalation phase, and fall represents the exhalation phase, respectively. Using the breathing data, we computed a synchronicity value for each interval or stroke (see the highlighted points in Figure~\ref{fig:synchronicityCalculation}). A peak detection library~\cite{peakdet2019} identified local extrema. Subsequently, the deviation of the breathing maximum from the ideal center between two neighboring rowing maxima was computed. This deviation, represented as a percentage value, encodes the breathing synchronicity of one rowing cycle. It is optimal when the inhalation peaks while the handlebar arrives at the closest point to the machine. Figure~\ref{fig:synchronicityCalculation} provides some examples of the synchronicity computation. For the statistical analysis, all the synchronicity values within one condition were averaged, resulting in one value per condition and participant.

\subsection{Procedure}
Every participant started with condition one (see Table~\ref{tab:conditions}) in which rowing strokes and breathing rhythm were recorded without any instructions regarding a correct breathing pattern. This first conditions acted like a naive, non-informed baseline. Subsequently, the haptic impulse actuator belt was set up around participants' lower back, and they were informed about the haptic and visual feedback which they would receive in the following conditions. Also, they were instructed on the correct breathing patterns for rowing (inhale in the recovery phase, exhale in the drive phase). Then conditions two to seven were run in a randomized order. After each condition, participants were asked to fill out two questionnaires on a tablet: Participants' perception of flow (full immersion, feeling of focus and involvement in an activity) was measured using the Flow Short Scale \cite{jackson2010flow} (discrete 5-point Likert scale). The AttrakDiff Mini \cite{hassenzahl2003attratctivediff} (discrete 7-point Likert scale) was used in order to measure differences in received attractiveness of the different conditions. It comprehends four dimensions: pragmatic quality (perceived usability), hedonic quality - stimulation (perceived novelty and interestingness), hedonic quality - identity (perceived identification with the product), and attractiveness (perceived quality). Before each VR condition, participants were asked to put on the HMD. At the end of the experiment, participants rated which feedback modality they liked best and which they found the most helpful by ordering photos from best to worst.

\subsection{Experimental Design}
Table~\ref{tab:conditions} outlines the set of conditions used. It is important to highlight that condition one and two differ in the instructions provided. Visual feedback was only provided through the VR system, which is why the design is not fully factorial.

\begin{table}[htb]
\centering
\begin{tabular}{cccc}
    \toprule
    Condition & Virtual Reality & Visual Feedback & Haptic Feedback \\ 
    \midrule
    1 (baseline) & \cellcolor{sonialightgray}No & \cellcolor{sonialightgray}No & \cellcolor{sonialightgray}No  \\
    2 & \cellcolor{sonialightgray}No & \cellcolor{sonialightgray}No & \cellcolor{sonialightgray}No \\
    3 & Yes & \cellcolor{sonialightgray}No & \cellcolor{sonialightgray}No \\
    4 & Yes & Yes & \cellcolor{sonialightgray}No \\
    5 & Yes & Yes & Yes \\
    6 & Yes & \cellcolor{sonialightgray}No & Yes \\
    7 & \cellcolor{sonialightgray}No & \cellcolor{sonialightgray}No & Yes \\
    \toprule
\end{tabular}
\caption{Experimental conditions comparing the two different feedback modalities in VR and without VR. All participants started with condition 1, the order of the remaining conditions was randomized.}
\label{tab:conditions}
\end{table}

\section{RESULTS}
\subsection{Instruction and the VR Environment}
A Wilcoxon signed-rank test indicated that there was a significant influence of verbal instruction on the synchronicity ($V=95$, $p<.001$). For this, the synchronicity was compared between the baseline (condition 1, without instruction) and condition 2 (with instruction). See Table~\ref{tab:descriptive_results} for details.

Assessing the impact of the virtual environment on the BMS, data from condition two and three, both without any feedback modalities but with and without the virtual environment was compared. On a descriptive level, the mean synchronicity without VR and without any kind of feedback (condition 2) was 57.392\% ($SD=31.9\%$), while using VR gained 57.442\% ($SD=31.9\%$, condition 3). Since the data was not normally distributed, a Wilcoxon signed-rank test was conducted, showing no significant differences between the two groups ($V=301$, $p=.5$). Therefore, the hypothesis regarding a difference purely based on the VR environment is not supported.

\subsection{Effects on Synchronicity}
A repeated measure analysis of variance (ANOVA) was employed to compare the two independent variables (IV) haptic and visual feedback on their impact on the dependent variable (DV) BMS. Both IVs included two levels each, using the feedback modality or not. A random factor for the participants was taken into account. On a descriptive level using only visual feedback increased the synchronicity by 2.911\% while haptic feedback gained 7.123\%. Both modalities combined yielded an increase of 8.839\% compared to no feedback at all (compare Table~\ref{tab:descriptive_results}).

\begin{figure}[htb]
  \centering
  \includegraphics[width=8.8cm]{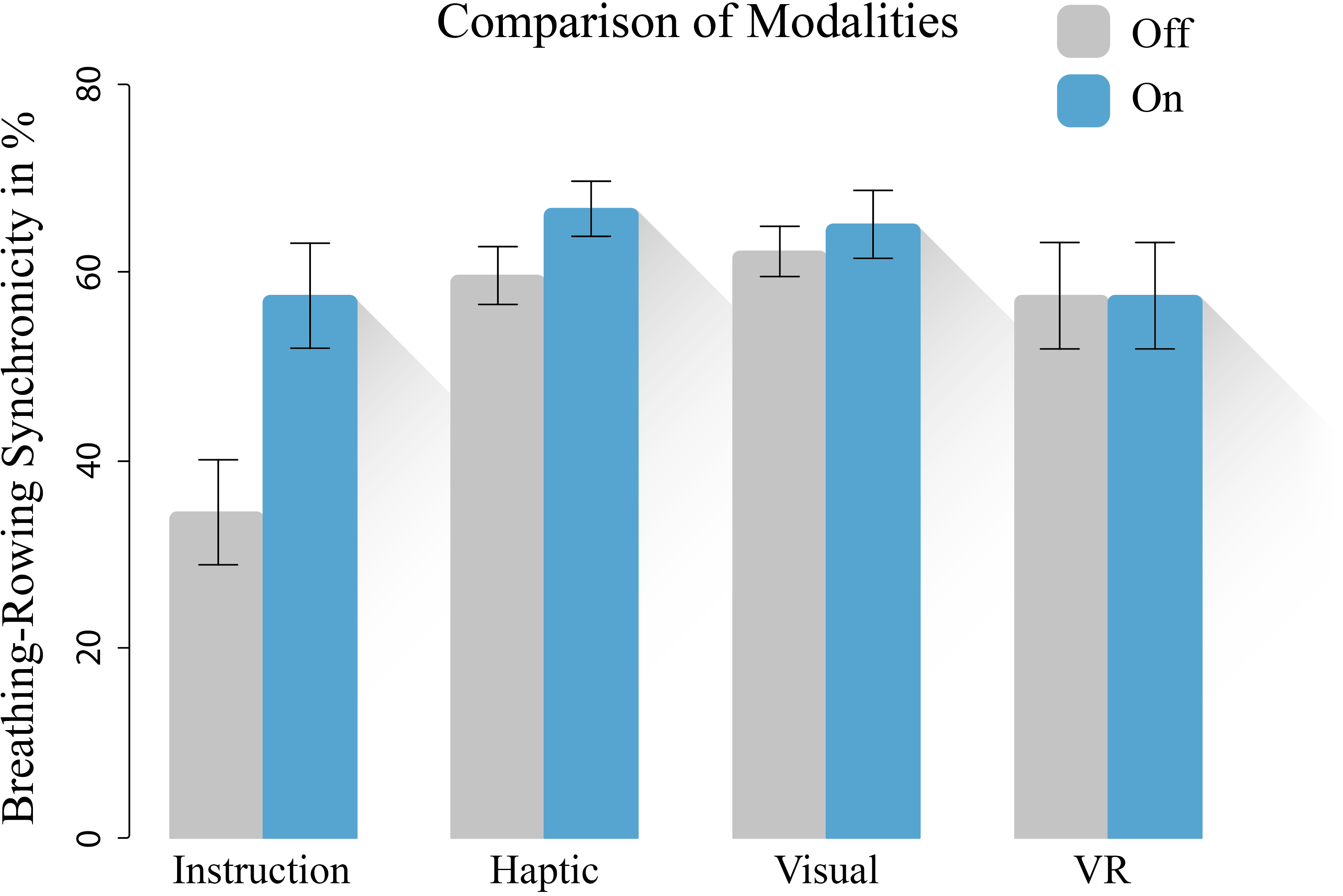}\\
  \caption{Bar plot comparing the impact of instruction, the two main factors haptic and visual feedback, as well as the impact of the VR system on the mean synchronicity. The difference between the two levels of instruction and haptic feedback are both statistically significant, while the visual feedback and the impact of VR are not. Whiskers depict the SEM.}
  \label{fig:mainEffects}
\end{figure}

A significant main effect of the haptic feedback ($F(1,31)=7.122$, $p=.012$), no significant main effect for the visual feedback ($F(1,31)=1.915$, $p=.176$) and a significant interaction effect between haptic and visual feedback ($F(1,31)=4.164$, $p=.049$) were found (see Figure~\ref{fig:mainEffects}). These results support the hypothesis that haptic feedback does increase the synchronicity, whereas there is no evidence for a positive influence of the visual feedback.


\begin{table}[]
    \centering
    \begin{tabular}{cccccc}
        \toprule
        Variable & N & Mean & SD & 95\% CI & signif.\\
        \midrule
        naive & 32 & 34.534 & 30.894 & 23.396 - 45.673 & \multirow{2}{*}{$p \leq.001$}\\
        instructed & 32 & 57.393 & 31.9 & 45.892 - 68.894 &\\
        \addlinespace
        haptic on & 96 & 66.706 & 28.731 & 60.884 - 72.527 & \multirow{2}{*}{$p \leq.05$}\\
        haptic off & 96 & 59.582 & 30.411 & 53.421 - 65.744 &\\
        \addlinespace
        visual on & 64 & 65.085 & 27.97 & 58.098 - 72.072 & \multirow{ 2}{*}{\textemdash}\\
        visual off & 128 & 62.174 & 30.619 & 56.818 - 67.529 &\\
        \addlinespace
        VR on & 32 & 57.393 & 31.9 & 45.892 - 68.894 & \multirow{ 2}{*}{\textemdash}\\
        VR off & 32 & 57.442 & 31.889 & 45.945 - 68.939 &\\
        \toprule
        \multicolumn{6}{l}{$\mathit{N}=\text{observations}$, $\mathit{SD}=\text{standard deviation}$, $\mathit{CI}=\text{confidence interval}$}
    \end{tabular}
    \caption{Impact on BMS; variables \textit{naive} and \textit{instructed} describe the comparisons of condition 1 and 2 only, the initial one without oral instructions on how to breath correctly during rowing and a comparison later on in the experiment with no assistance but after the instructions. Likewise, the impact of the \textit{VR} system was investigated by comparing condition 2 and 3 only. The variables haptic and visual depict the grand mean of conditions 2 to 7 combined, equally to the data used for the ANOVA. Hence, the difference of observation counts.}
    \label{tab:descriptive_results}
\end{table}

\subsection{Effects on UX}
\begin{figure}[htb]
  \centering
  \includegraphics[width=8.8cm]{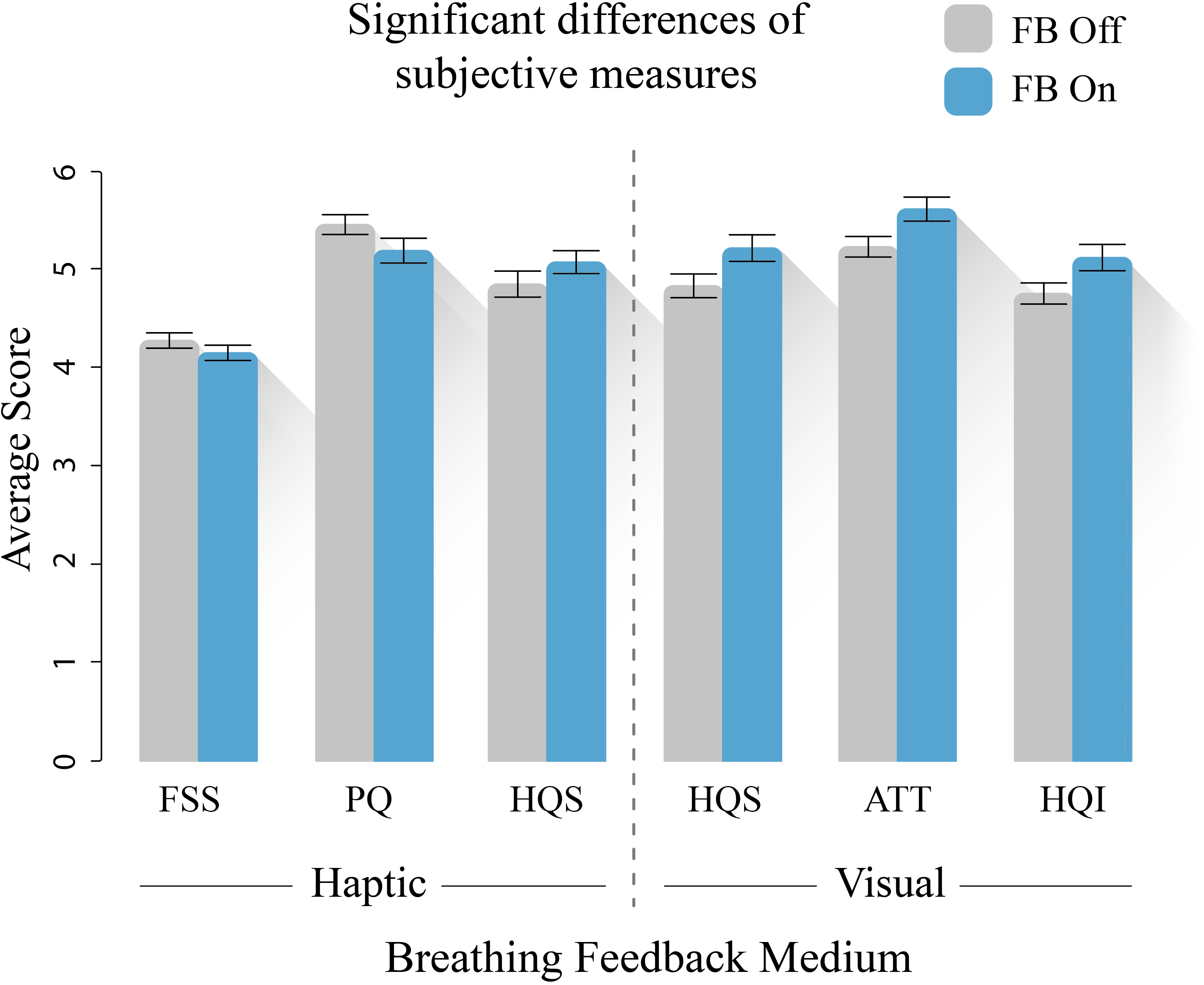}\\
  \caption{Statistically significant differences of the two feedback modalities on UX. FFS~=~Flow Short Scale, PQ~=~pragmatic quality, HQS~=~hedonic quality stimulation, ATT~=~attractiveness, HQI~=~hedonic quality identity. Whiskers depict the SEM.}
  \label{fig:subjectiveMeasuresBarPlot}
\end{figure}

Similarly, repeated measures ANOVAs with the same IVs but using the questionnaires' scores as the DV were conducted to study the subjective measurements (see Figure~\ref{fig:subjectiveMeasuresBarPlot}).

\textbf{Flow:} A significant main effect of haptic feedback ($F(1,31)=4.561$, $p=.041$) on the perception of flow was found. Using haptic feedback significantly decreased the perceived flow from $M=4.268$ ($SD=.692$) to $M=4.143$ ($SD=.682$). Visual feedback had no effect on participants' perception of flow ($F(1,31)=.015$, $p=.904$), with a change from $M=4.202$ ($SD=.673$) not using it to $M=4.212$ ($SD=.723$) using visual feedback.

\textbf{Pragmatic Quality:} Concerning pragmatic quality, a significant main effect of haptic feedback was found ($F(1,31)=5.873$, $p=.021$). Similarly to the flow ratings, participants preferred conditions without the feedback modality. Ratings decreased from $M=5.448$ ($SD=1.024$) not using haptic feedback to $M=4.212$ ($SD=.723$) when using it. Visual feedback had no significant impact ($F(1,31)=2.577$, $p=.119$), changing the ratings from $M=5.258$ ($SD=1.19$) not using it to $M=5.438$ ($SD=.98$) with the modality.

\textbf{Hedonic Quality - Stimulation:} Both, haptic feedback ($F(1,31)=5.642$, $p=.024$) and visual feedback ($F(1,31)=11.38$, $p=.002$) have had a significant, positive effect on the stimulation aspect of the hedonic quality dimension. Haptic feedback increased the ratings from $M=4.849$ ($SD=1.234$) to $M=5.068$ ($SD=1.142$), visual feedback increased the ratings from $M=4.832$ ($SD=1.243$) to $M=5.211$ ($SD=1.042$).

\textbf{Hedonic Quality - Identity:} Regarding the identity aspect of hedonic quality, a significant main effect of the visual feedback was found ($F(1,31)=13.5$, $p<.001$; mean values changed from $M=4.758$, $SD=1.081$ without to $M=5.117$, $SD=1.034$ with visual feedback). Furthermore, the statistical analysis revealed an interaction effect between haptic and visual feedback ($F(1,31)=5.001$, $p=.033$). However, haptic feedback alone did not significantly alter the ratings ($F(1,31)=.046$, $p=.832$; mean values changed from $M=4.87$, $SD=1.086$ without to $M=4.885$, $SD=1.072$ with haptic feedback).

\textbf{Attractiveness:} The haptic feedback modality did not cause a significant difference in the attractiveness ratings ($F(1,31)=3.135$, $p=.08$), decreasing the ratings slightly from $M=5.427$, $SD=0.968$ without haptic feedback to $M=5.291$, $SD=1.0122$ with haptic feedback. However, the visual condition did impose a significant difference on the ratings ($F(1,31)=14.96$, $p<.001$) and improved them from $M=5.23$, $SD=1.019$ without visual feedback to $M=5.617$, $SD=0.881$.

For assessing the preference of participants for the three different interfaces (haptic, visual, and both), Friedman's Two-Way Analysis was conducted for the two variables likability rank and helpfulness rank between the interfaces in VR.

For both rankings likability ($\chi^2(2)=29.74$, $p<.001$, \& $df=3$, $N=32$) and helpfulness ($\chi^2(2)=20.51$, $p<.001$, \& $df=3$, $N=32$) statistical analyses have shown significant differences. 
Additionally, in order to find out pairwise significance, the posthoc test was done. Results show that the interface with no feedback ($M=3.31$, $SD=1.06$) was rated significantly less likable compared to the condition with visual feedback ($M=1.66$, $SD=.86$) and in comparison to feedback with both haptic and visual ($M=2.22$, $SD=.94$). Further on, only visual feedback was liked significantly better by participants compared to only haptic feedback ($M=2.81$, $SD=.89$). Similarly, helpfulness of no feedback ($M=3.25$, $SD=1.02$) was ranked to be significantly lower compared to visual feedback ($M=1.91$, $SD=.96$) and compared to both haptic and visual feedback ($M=1.91$, $SD=.96$).

\section{DISCUSSION}
The first statistical test confirmed our hypothesis of a positive impact of instructions provided before using a rowing ergometer. Even a quick explanation drastically improved the BMS. This simple fact is a particularly valuable result since it emphasizes the importance of a simple introduction. This is especially true for low-end fitness centers where users do not get any guidance or training before a workout.

The main goal of the present work was to investigate the impact of the visual and haptic feedback modalities. Although the hypothesis of a main effect of visual feedback did not hold, a significant main effect of the haptic feedback confirmed the corresponding hypothesis. Therefore, using haptic impulses to improve BMS is a feasible measure, and as the comparison between two conditions both with haptic feedback but with and without VR demonstrated, this effect is independent of the (virtual) environment. This VR independence is somewhat surprising since previous, similar studies - although using different measurements - have found a significant difference in breath counts using a similar VR environment compared to not using it~\cite{arndt2018using}. Hence, we similarly expected influence on the breath that impacts the synchronicity. However, this independence of the environment opens up many possibilities of assisting novices to intermediately skilled rowers in both VR exergames as traditional workouts.

Augmenting audiovisual information with tactile stimuli can increase the perceived QoE, although the overall liking seems to be dependent on the users' expectations (e.g.,~\cite{maggioni2017measuring, giroux2019haptic}). Contrastingly, our participants did not favor the haptic feedback conditions (dimensions flow, and perceived usability). This might be an artifact of the haptic interface used rather than of the modality itself. The relatively large belt might be uncomfortable to wear while performing physically demanding rowing movements. Therefore, future work should consider different approaches to delivering tactile feedback.

Regarding the visual feedback, our data reveals some interesting discrepancies between subjective perception and objective measurement. Although our hypothesis of a positive impact of the visual feedback could not be statistically supported, it was the visual feedback modality that participants rated best. This is true for the rank task regarding helpfulness and preferability, but also attractiveness and both aspects of hedonic quality. Moreover, while the visual feedback did not score significantly higher on the pragmatic quality and flow scale, the opposite is given; the haptic condition was significantly downrated in these two dimensions. Consequently, the visual feedback was accepted better - without actually improving performance, while haptic feedback did indeed.

A surprising finding of this work concerns the significant interaction between the two feedback modalities. The synchronicity peaked when using the haptic modality alone; resulting in a reduction of performance when the haptic feedback was used simultaneously with the visual feedback (although not statistically significant). There are two possible explanations for this finding. First, the two stimuli are not only different in their modality, but also in their perception: one is a constant animation (visual) while the other is only a short impulse (haptic). This difference might render the two incomparable and should be investigated further. Second, the difference might be an artifact of the experimental design. Since the visual feedback was bound to the VR environment, the design is not fully crossed. Therefore, the two extrema of the interaction (no haptic but visual feedback as well as both together) are averages from only one condition each (no. 4 \& no. 5 respectively), resulting in 32 observations. Both other cases of the interaction consist of 64 observations. Thus, the cause of the interaction could be a result of our experimental design, which should be considered in future work.

\section{CONCLUSION}
The present work involved analyzing the effect of breathing feedback on the BMS. We contrasted haptic sensations and visual animations as feedback modalities together with or without a VR environment and further tested the impact of simple, verbal instruction. Breathing feedback was provided by analyzing the rowing movement pattern together with a high-frequency sampling of the participants' respiration.

All participants' mean synchronicity increased after receiving verbal guidance. The results indicated that the breathing feedback presented had a significant impact on the BMS when delivered via a haptic impulse. Interestingly, although our visual feedback did not increase the synchronicity, it was rated higher in almost all assessed UX dimensions and favored over the haptic feedback, which was partly even downgraded in the UX related ratings.

Improving the BMS helps athletes to enhance health and performance~\cite{ramacharaka2007science}. This work contributes to the body of research not only how beneficial a quick introduction can be, but also evidence for a positive impact of a pure feedback sensation on performance - with or without a VR environment. Further research should investigate the impact of an improved BMS on plain rowing performance in terms of speed and endurance. 



\bibliographystyle{IEEEtran}
\bibliography{references.bib} 

\end{document}